\documentclass[a4paper,11pt]{article}

\usepackage{pos}
\usepackage{lipsum} 
\usepackage{float}
\usepackage{subcaption}
\usepackage{caption}
\usepackage{wrapfig,graphicx}
\usepackage{amsmath}
\usepackage{tabularx}

\usepackage{pbox}
\usepackage{array}
\usepackage[export]{adjustbox}
\usepackage{natbib}
\usepackage{comment}
\newcolumntype{P}[1]{>{\centering\arraybackslash}p{#1}}

\title{Classification of high energy muon bundles and single muons from the southern sky  in IceCube}
\ShortTitle{classification of bundle vs single muons}

\author{The IceCube Collaboration \\{\normalsize \normalfont(a complete list of authors can be found at the end of the proceedings)}\\}

\emailAdd{moureen@udel.edu}
\emailAdd{dseckel@udel.edu}

\abstract{The IceCube Neutrino Observatory, located at the geographic South Pole, uses the glacial ice volume to detect astrophysical neutrinos. Detection of the neutrinos from the northern sky provides the opportunity to use a large effective volume. However, as the cross-section increases with energy, most high-energy neutrinos are absorbed by the Earth. On the other hand, probing down-going PeV neutrinos from the southern sky becomes challenging because of the large cosmic ray induced muon backgrounds. This contribution presents a method for classifying atmospheric muon bundles and single muons by analyzing the lateral and longitudinal characteristics of through-going track-like events from the southern sky. Muons generated in cosmic ray air showers form muon bundles, exhibiting a lateral spread spanning tens of meters within IceCube. We explore the time residual feature for the observed Cherenkov light to separate bundles from single muons. We also utilize energy losses along the track, which lead to fluctuations in the light intensity: bundles follow a smooth pattern, whereas single muons are more stochastic. A Boosted Decision Tree algorithm is trained on simulated, well-reconstructed cosmic ray and neutrino events to classify neutrino-induced single muons and cosmic ray bundles.

\vspace{4mm}

{\bfseries Corresponding authors:}
Najia Moureen Binte Amin$^{1}$,
David Seckel$^{1*}$,\\
{$^{1}$ \itshape Bartol Research Institute, Department of Physics and Astronomy, University of Delaware}\\
$^*$ Presenter

}

\ConferenceLogo{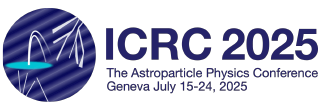}

\FullConference{39th International Cosmic Ray Conference (ICRC2025)\\
 15–24 July 2025\\
Geneva, Switzerland\\}

\begin{document}

\maketitle

\section{Introduction}\label{sec1}
The IceCube Neutrino Observatory is embedded within a cubic kilometer of Antarctic ice. IceCube’s unique design allows it to detect neutrinos from all directions, making it especially well suited for all-sky searches, though its sensitivity varies with direction.  The IceCube Observatory has two components: the in-ice array~\cite{collaboration_icecube} and IceTop surface array~\cite{collaboration_icetop_2013}. The in-ice array is located at depths between $1450$~m and $2450$~m. This array has $ 86$~strings set up in a hexagon-shaped pattern with an inter-string distance of $125$~m. Each string is equipped with $60$~Digital Optical Modules (DOMs) spaced 17 meters apart vertically, which detect Cherenkov light from charged particles exceeding the speed of light in ice.  IceTop, positioned on the surface above the in-ice array, includes $ 162$~ice Cherenkov tanks grouped in pairs of $81$~stations.

\begin{figure}[b]
  \begin{subfigure}[b]{0.5\textwidth}
    \includegraphics[width=\textwidth]{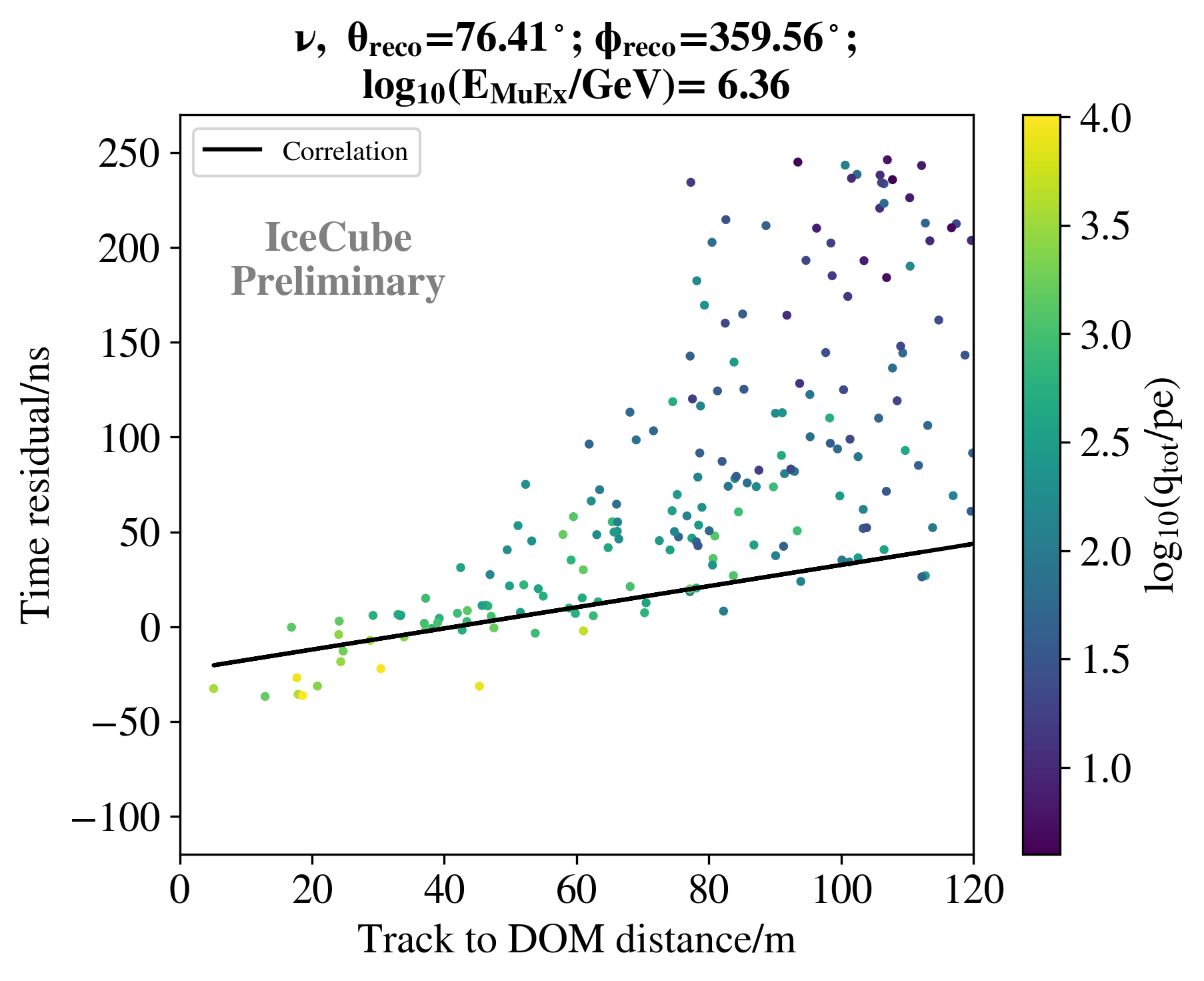}
  \end{subfigure}
  \hfill
  \begin{subfigure}[b]{0.5\textwidth}
    \includegraphics[width=\textwidth]{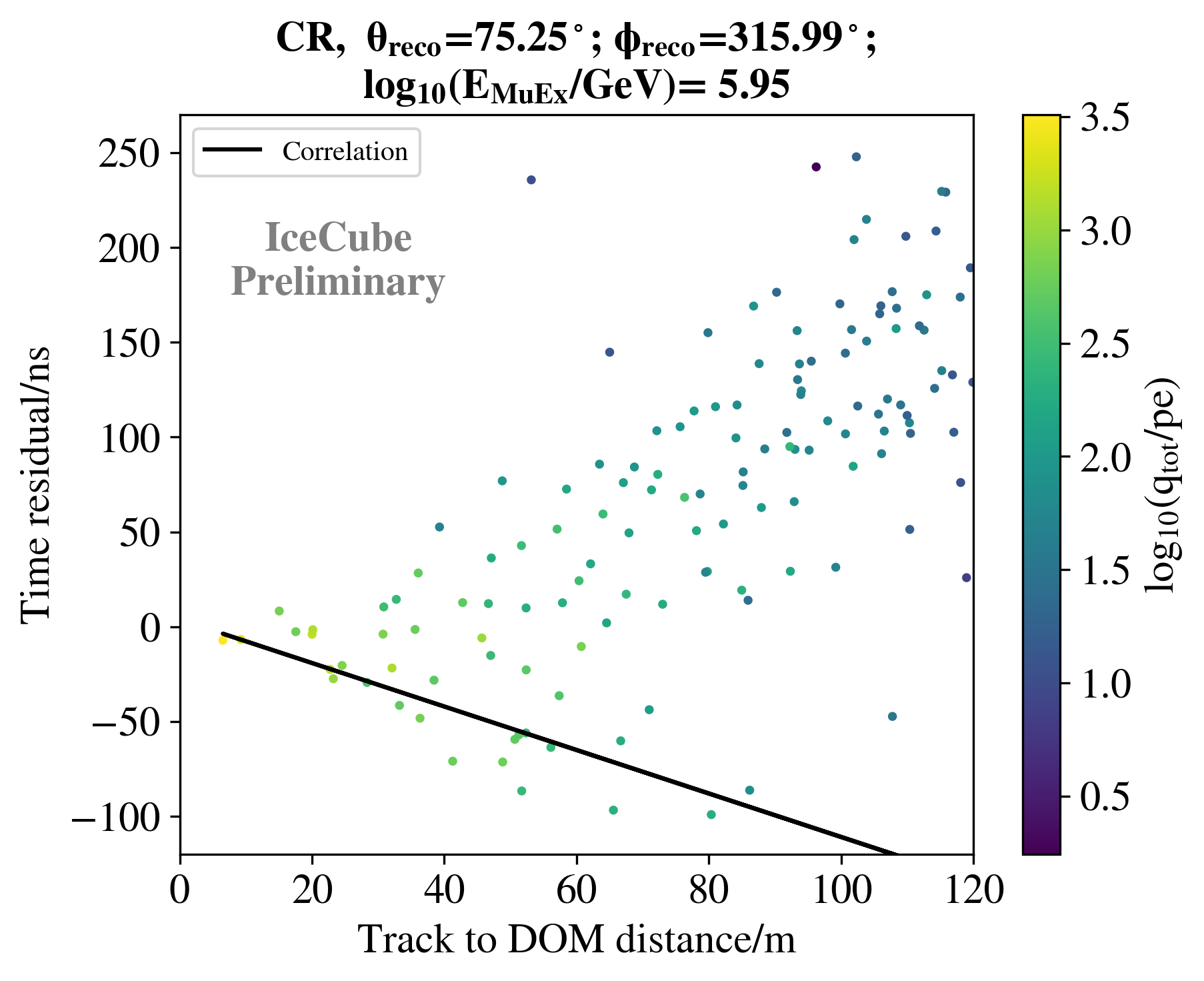}
  \end{subfigure}
  \caption{Example events illustrating a neutrino-induced track(left) and a cosmic ray muon bundle (right). The x-axis denotes the perpendicular distance from the reconstructed track to the DOM, while the y-axis shows the time residual of the earliest pulse in each DOM  (i.e., the difference between the observed pulse time and the expected arrival time of unscattered Cherenkov photons). The color scale represents the logarithm of the total charge recorded in each DOM.  Neutrino events typically exhibit a flat and compact time residual pattern very near the track, consistent with single-track light emission. On the other hand, cosmic-ray events show pulses with early time residuals due to multiple muons arriving with spatial differences resulting in temporal offsets in photon arrival at the DOMs. The black line shows correlation between time residual and track to DOM distance which is covered in detail in section~\ref{sec3}.
 }
 \vspace{-0.5\baselineskip}
\label{fig_nu_cos}
\end{figure}

\begin{figure}
\centering
  \includegraphics[width=1\textwidth]{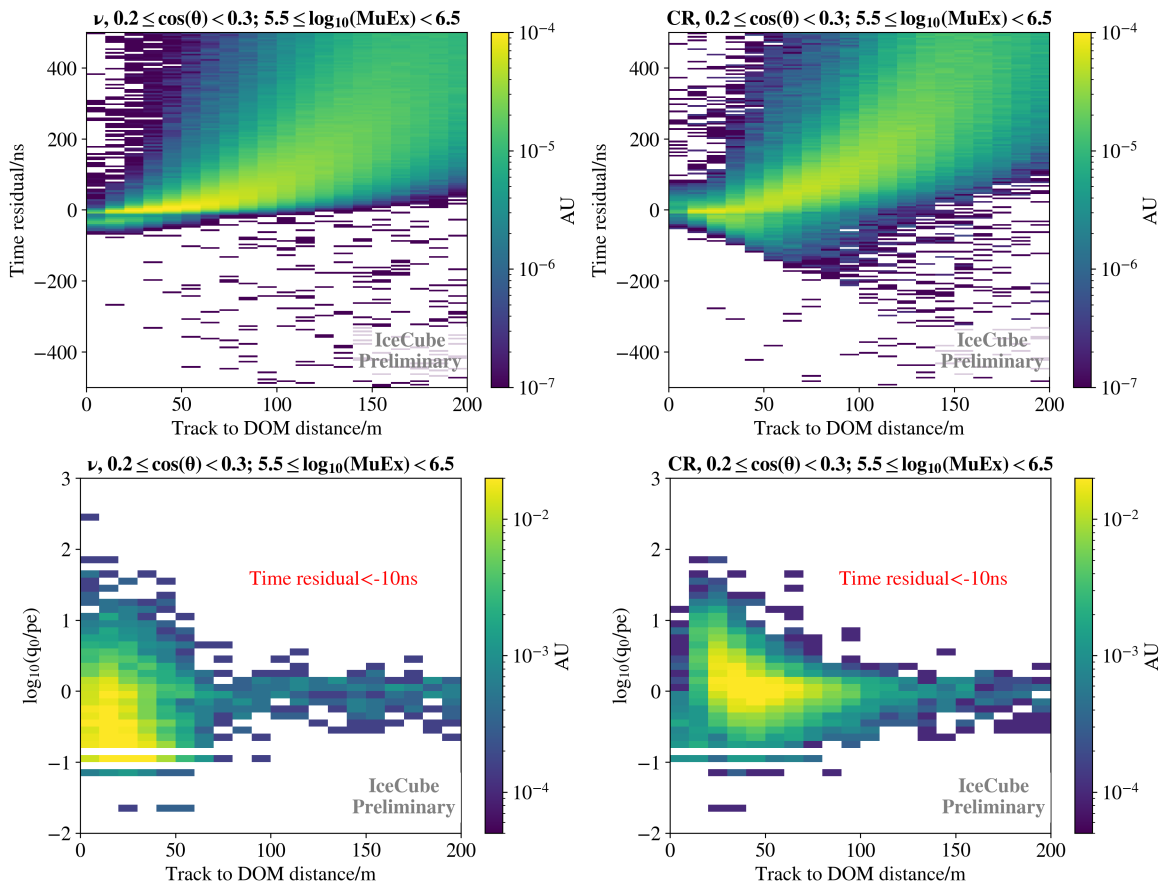}  

  \caption{Two-dimensional histograms of neutrino and cosmic-ray events passing the GFU filter. The top panels present time residuals as a function of perpendicular distance from the reconstructed track. Neutrino-induced events (top left) display a flat and localized distribution near the track, and cosmic-ray-induced muon bundles (top right) exhibit prominent early time residual features. These observed features are consistent with Fig.~\ref{fig_nu_cos}. In the neutrino distribution, there is an excess of early pulses that distinctly peaks at $-30~$ns attributed to prepulses~\cite{collaboration_calibration_2010} rather than parallel muons. The bottom panels highlight the differing origins of early light between the two event classes, isolating pulses with time residuals less than $-10~$ns. These 2D histograms show the earliest pulses as a function of track-to-DOM distance and first deposited charge in each DOM. The neutrino sample (bottom left) shows a broader distribution with lower first-hit charges in DOMs closer to the track, likely associated with prepulses.  In the cosmic-ray sample(bottom right), early pulses cluster beyond $\sim20~m$ from the track with moderate charges, consistent with early hits from muon bundles. Additionally, DOMs that record prepulses often carry higher total charge compared to the early pulses because of parallel muons, which further contributes to the discrimination. The window shown in these plots differs from that in Fig.~\ref{fig_nu_cos}, and is used for the $\Delta\log\mathcal{L}$ variable, which will be discussed in section~\ref{sec3}.}
  \label{fig_nu_cos_2}
    \vspace{-1\baselineskip}
\end{figure}

IceCube detects high-energy astrophysical neutrino candidates and sends alerts within a minute to the multimessenger community to identify sources of these high-energy particles through the coincident detection of transient events across multiple messengers. The first evidence of high-energy astrophysical neutrinos in IceCube came from events that started inside the detector~\cite{science_2013}.  Later, its sensitivity was extended by studying extended muon tracks traversing the detector from the northern sky, where the Earth shields the detector from the atmospheric muon background.
While below $100~$TeV, the atmospheric neutrinos dominate, above a PeV, the Earth becomes opaque to neutrinos due to their increasing interaction cross section. This results in an upward window of $100-1000$~TeV. To probe PeV neutrinos, IceCube focuses on downward-going events from the southern hemisphere. The primary background for down-going high-energy searches consists of muon bundles generated by cosmic ray air showers in the Earth’s atmosphere, while neutrino interactions are expected to produce single muons. Previous analyses~\cite{collaboration_realtime_2017} implemented a stringent zenith dependent cut on charge as an alert selection criteria to suppress the dominant atmospheric lepton background.  In this work, we investigate lateral and longitudinal features of cosmic ray and neutrinos to distinguish between signals and backgrounds, using a Boosted Decision Tree (BDT) classifier. Also, we focus on relaxing the strict zenith and charge-dependent cuts for the southern sky, enabling more astrophysical neutrinos without increasing the background, which yields a signal-to-noise ratio of $\sim11$.

\section{Simulation and Event selection}\label{sec2}
In this analysis, we use only simulated events to train the BDT. Atmospheric muon backgrounds are generated using the CORSIKA air shower simulation framework~\cite{heck_extensive_2001} with the SIBYLL 2.3~\cite{Sybill2.3} hadronic interaction model, covering primary energies from $10^6~$ to $10^{10}~$ GeV. The cosmic ray background is weighted using the GaisserH4A model~\cite{GaisserH4A}. Muon neutrino events are simulated using a neutrino-generator~\cite{nugen}, with primary energies ranging from $10^5~$GeV up to $10^8~$GeV. We model the astrophysical neutrino component following the spectral shape and normalization reported in earlier IceCube results~\cite{abbasi_improved_2022}. The Spice3.2~\cite{rongen_calibration_nodate} ice model is used in the simulations to describe the optical properties of the South Pole ice.

The analysis begins with a sample of high-quality, track-like gamma-ray follow-up(GFU) candidates~\cite{kintscher_rapid_2020}  from the southern sky ($0.2\mathrm{\leq \cos{\theta}<}1$) with a rate of $\sim 2$mHz. GFU selection identifies potential astrophysical neutrinos based on BDTs. Only the most energetic events from the GFU selection are shared as alerts with the multimessenger community to ensure a high-purity.

The MuEx~\cite{collaboration_energy_2014} energy proxy is used in this analysis which estimates the muon’s energy based on the Cherenkov light produced by its energy losses along the track. In this model, the muon is approximated to emit Cherenkov light uniformly along its path, with photon detection varying by distance due to scattering close by and absorption at greater distances, averaged over the optical properties of ice layers. These two ranges are linked by an empirical formula. Events are selected within the energy proxy range $5.5\leq \log_{10}(\mathrm{MuEx})<6.5$. This analysis uses the SplineMPE~\cite{schatto_stacked_2014} reconstruction, a maximum likelihood method designed to reconstruct muon tracks in IceCube. It fits the observed photon arrival times using pre-calculated splines that account for detailed light propagation through the ice. We also require that the reconstructed muon track has a minimum length of $200~$m within the detector volume to achieve a reliable reconstruction quality. 

Moreover, the selection criteria are designed to retain events with well-sampled lateral light distributions. To achieve this, three additional requirements are applied.  We select events with at least three DOMs which fall within a spatial and temporal window relative to the reconstructed track and whose earliest pulse charge satisfies $\log_{10}Q_{0} >- 0.5$. The spatial range $\Delta x$ is defined by the distance to the DOM with the minimum time residual, or set to $50~$m if smaller.  The temporal window includes DOMs with time residuals between the minimum and $50~$ns above the time residual of the DOM closest to the track. For this specific selection criterion, we restrict DOMs to those within $200~$m of the track and with time residuals larger than $- 200~$ns, in order to suppress the influence of random noise.

\section{Key Variables for BDT-based Classification}\label{sec3}

\begin{figure}[h]
  \begin{subfigure}[b]{0.4\textwidth}
    \includegraphics[width=\textwidth]{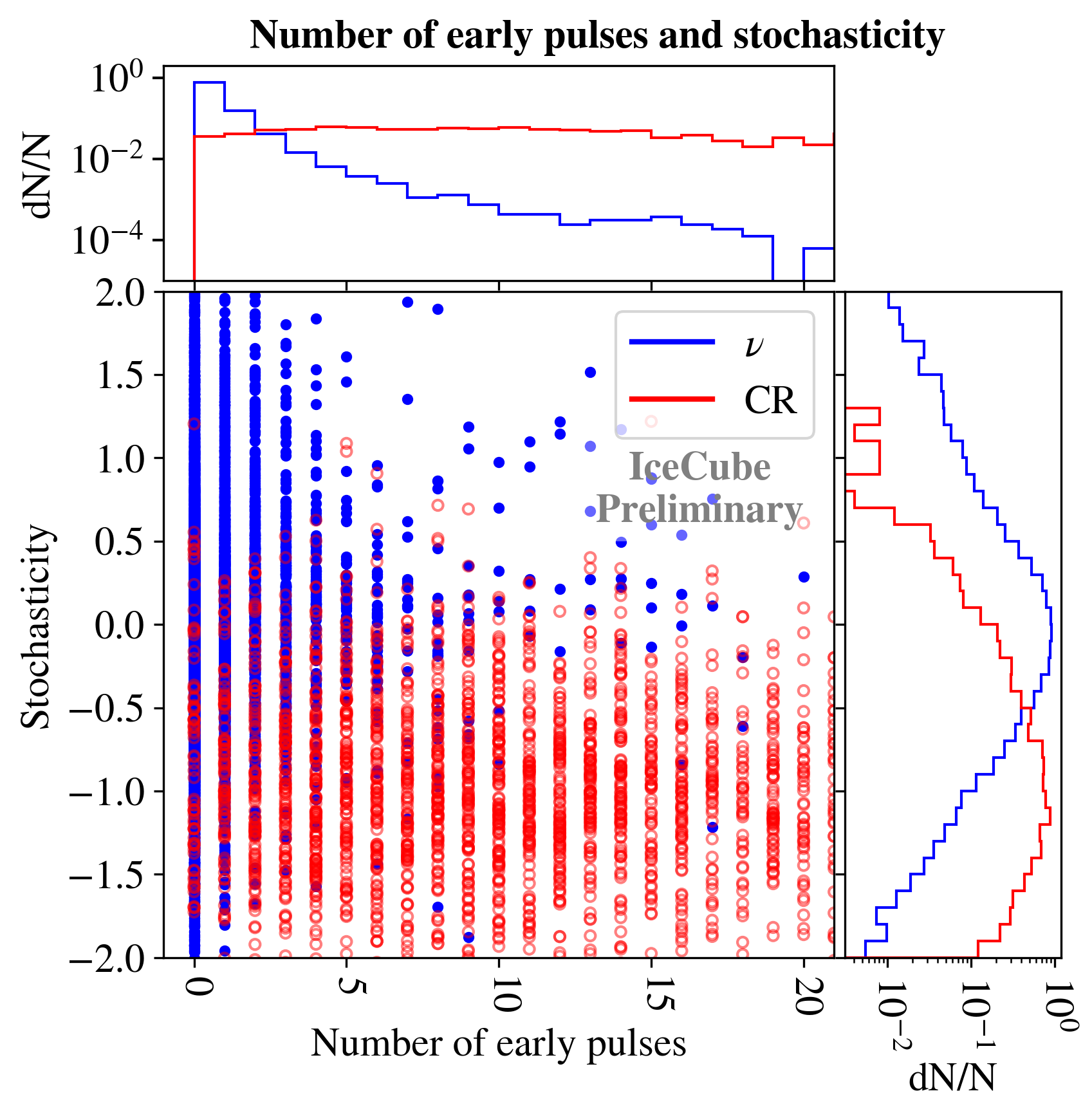}
  \end{subfigure}
  \centering
  \hspace{1em}%
  \begin{subfigure}[b]{0.4\textwidth}
    \includegraphics[width=\textwidth]{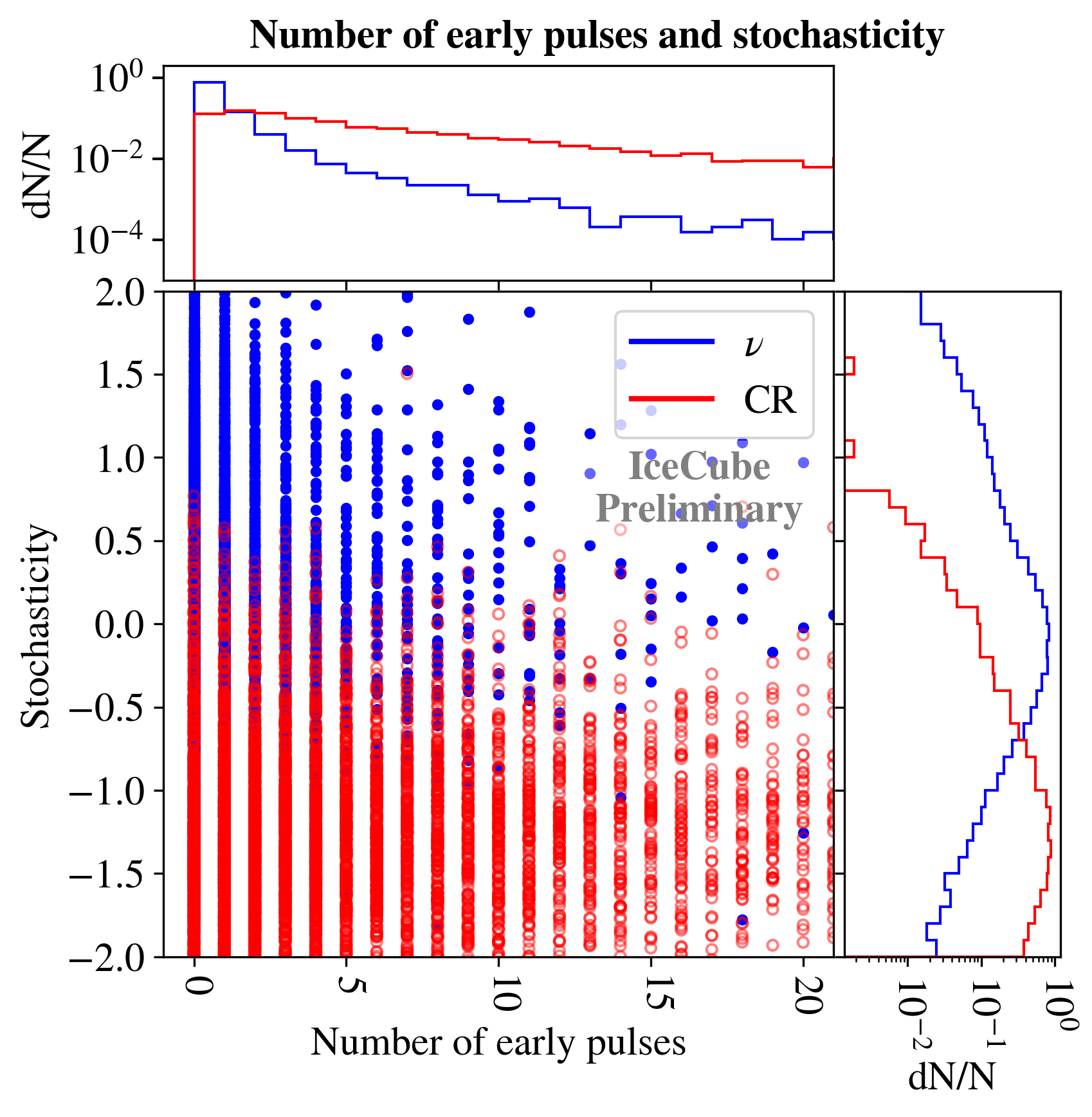}
  \end{subfigure}
  \centering
  \hspace{1em}%
  \begin{subfigure}[b]{0.4\textwidth}
    \includegraphics[width=\textwidth]{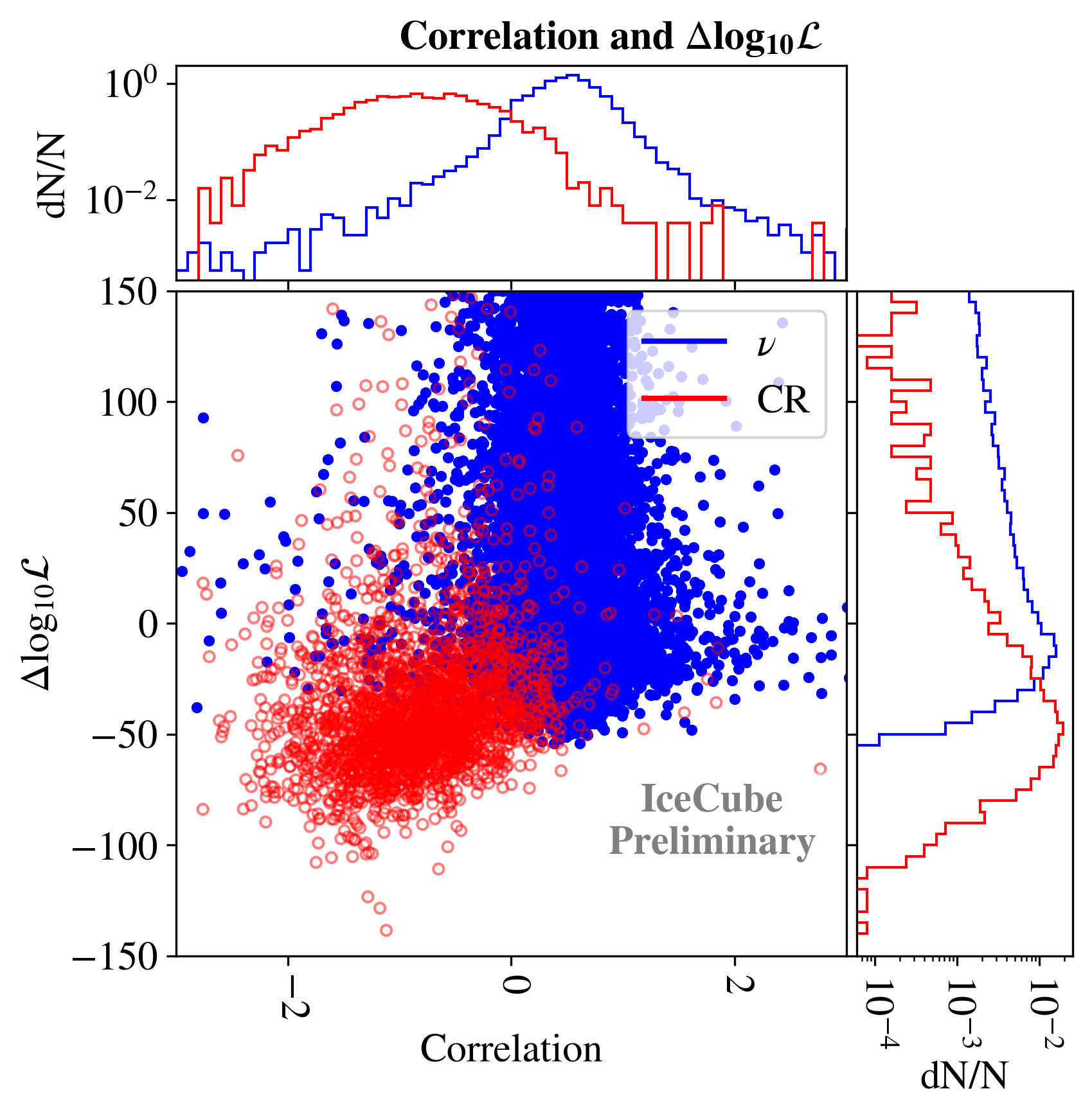}
  \end{subfigure}
   \centering
  \hspace{1em}%
  \begin{subfigure}[b]{0.4\textwidth}
    \includegraphics[width=\textwidth]{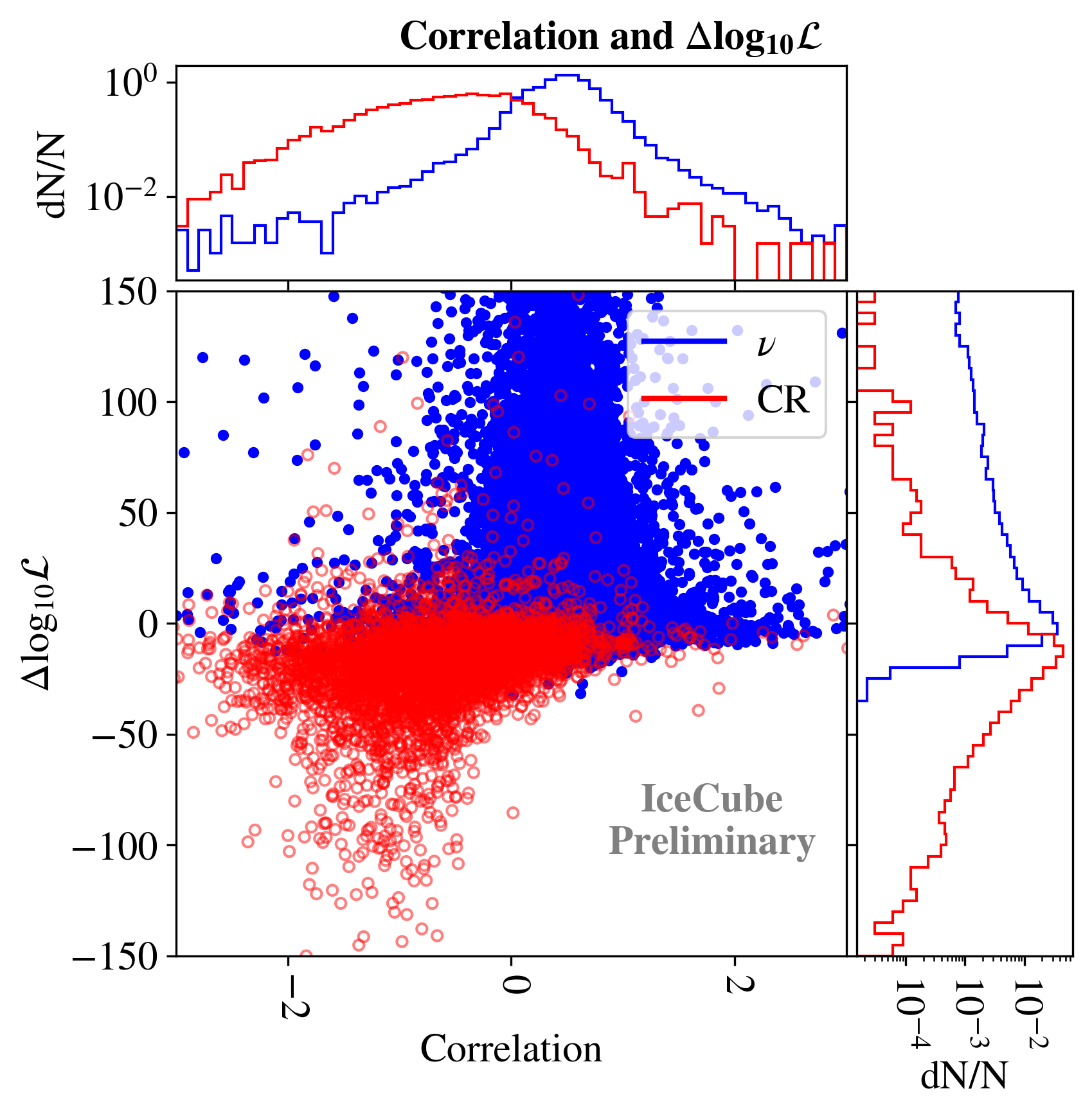}
  \end{subfigure}
 
  \caption{ Distributions of reconstruction-based input variables for cosmic-ray (red) and neutrino (blue) events, illustrating their separation power for horizontal (left, $0.2\leq \cos{\theta}<0.3$) and vertical (right, $0.7\leq\cos{\theta}<0.8$) events. Lateral features show improved separation in horizontal events due to the $17~$m DOM spacing.
}
  \label{hist_all}
\end{figure}

High-energy cosmic rays initiate extensive air showers that produce bundles of nearly parallel muons, causing lateral offsets and early light signatures in the DOMs. On the other hand, neutrino events exhibit a uniform time residual pattern near the track, consistent with single-muon light emission, with increasing delays at larger distances due to scattering. This contrast can be exploited to distinguish bundles from single muons. Fig.~\ref{fig_nu_cos} and~\ref{fig_nu_cos_2} illustrate examples of both neutrino-induced muons and cosmic-ray muon bundles. Additionally, using the total deposited charge and the charge of the earliest hit in DOMs further enhances the separation as shown in Fig.~\ref{fig_nu_cos_2}. We also incorporate the stochastic energy loss along the track as one of the input variables for the classification. Fig.~\ref{hist_all} illustrates how the parameters separate neutrinos and cosmic rays for both vertical and horizontal events and also shows that the variables are largely independent. The details of the key variables used to capture these features are described in the following paragraphs.

\textbf{Number of early pulses:}
To quantify the number of early pulses, we count the DOMs with time residuals smaller than the minimum time residual among the three DOMs closest to the reconstructed track within the spatial, temporal, and charge selection defined in section~\ref{sec2}. To mitigate the contribution from prepulses, we apply a charge threshold requiring the first pulse in each DOM to satisfy $\log_{10}(Q_0) > -0.5$, as described in section~\ref{sec2}. Although the applied charge cut helps suppress prepulses, some may persist. To further reduce their contribution, we only count early pulses for this variable only from DOMs located at least $20~$m from the track, where contributions from early light due to parallel muons are more prominent and prepulse contamination is less likely to dominate.

\textbf{Correlation:}
Following the same selection scheme detailed in section~\ref{sec2}, but using only DOMs with time residuals below the median of the defined window, we fit the time residuals as a function of the track-to-DOM distance using early pulses, as shown in Fig.~\ref{fig_nu_cos}. In cosmic-ray events, early pulses from the muon bundle tend to exhibit a decreasing time residual with increasing distance from the track. In contrast, neutrino events typically show an increasing time residual trend with distance due to scattering. We define the slope of this fit as the correlation, as shown in Fig.~\ref{fig_nu_cos}.

\textbf{Stochasticity:}
High-energy neutrino-induced muons lose energy stochastically, while muon bundles of equal energy distribute it among many muons, resulting in smoother energy loss. In the earlier study~\cite{yang_probing_2025}, stochasticity was found to increase strongly with energy, even in the case of large muon bundles. Unlike the previous approach, we fit the differences in energy loss between successive $120~$m bins, rather than the absolute loss values. The weighting is applied using the uncertainty, $\sigma = \sqrt{\Delta (dE/dx)}$. 

\begin{figure}[t]
  \begin{subfigure}[b]{0.32\textwidth}
    \includegraphics[width=\textwidth]{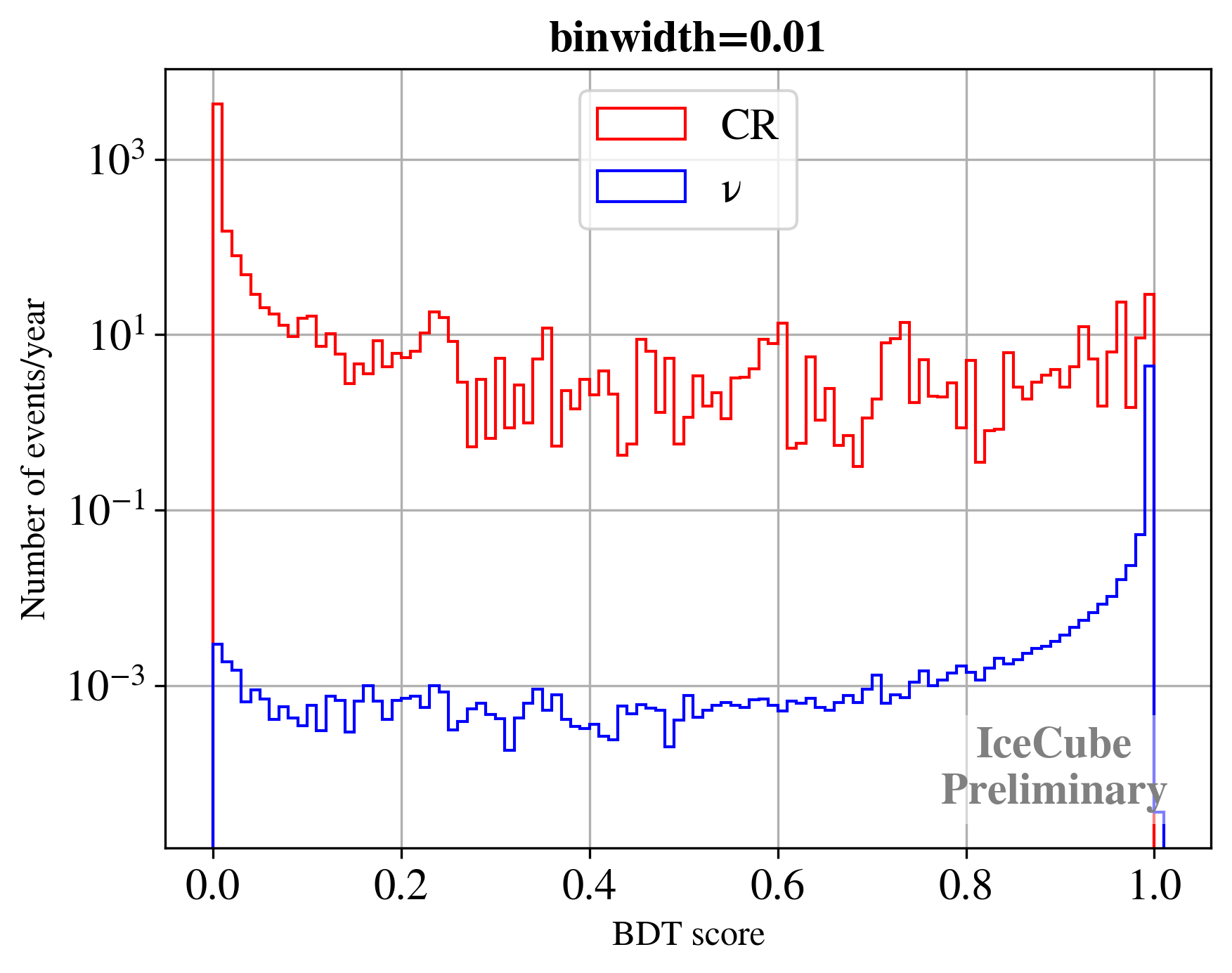}
    \end{subfigure}
    \hfill
  \begin{subfigure}[b]{0.32\textwidth}
    \includegraphics[width=\textwidth]{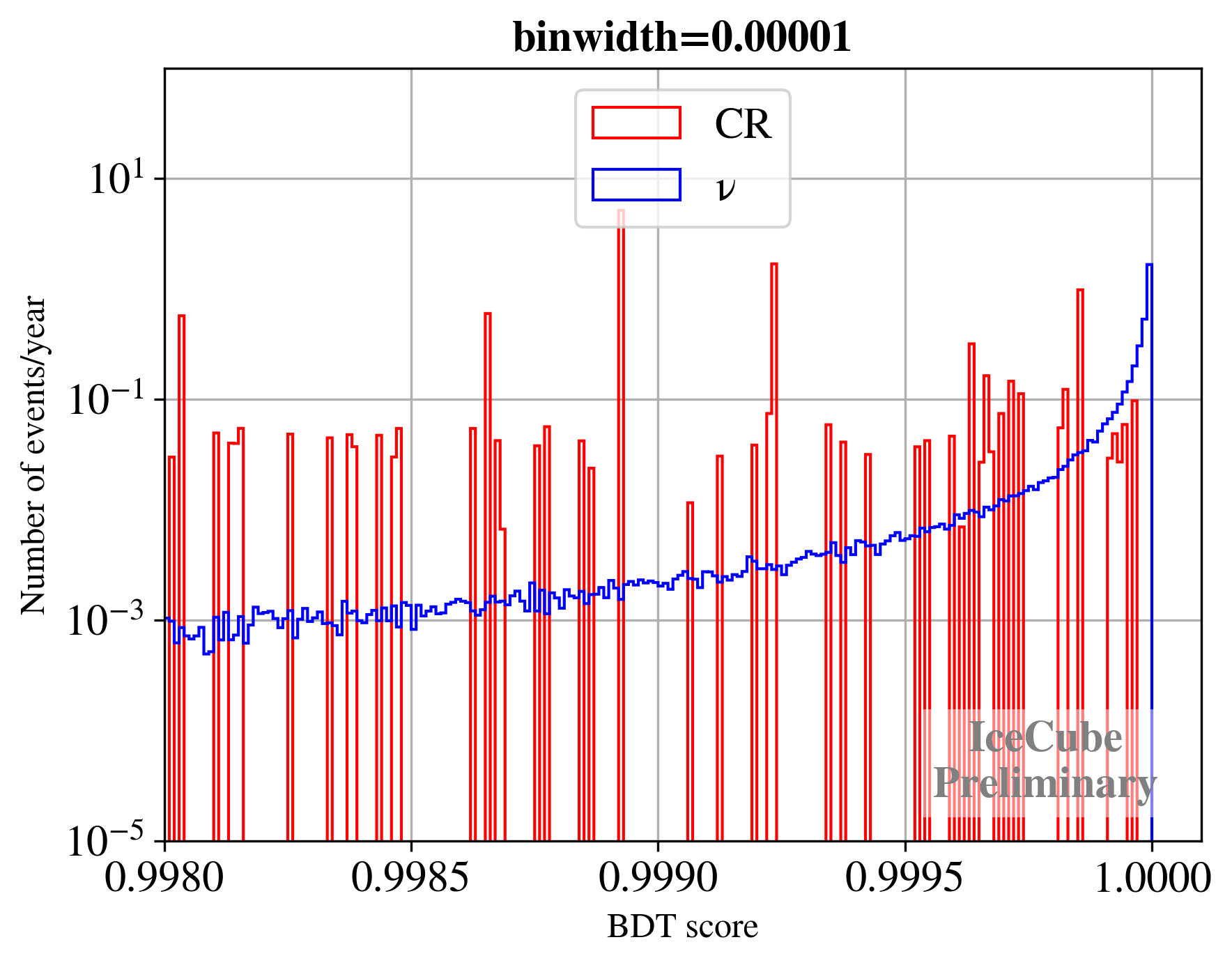}
    \end{subfigure}
    \hfill
    \begin{subfigure}[b]{0.32\textwidth}
    \includegraphics[width=\textwidth]{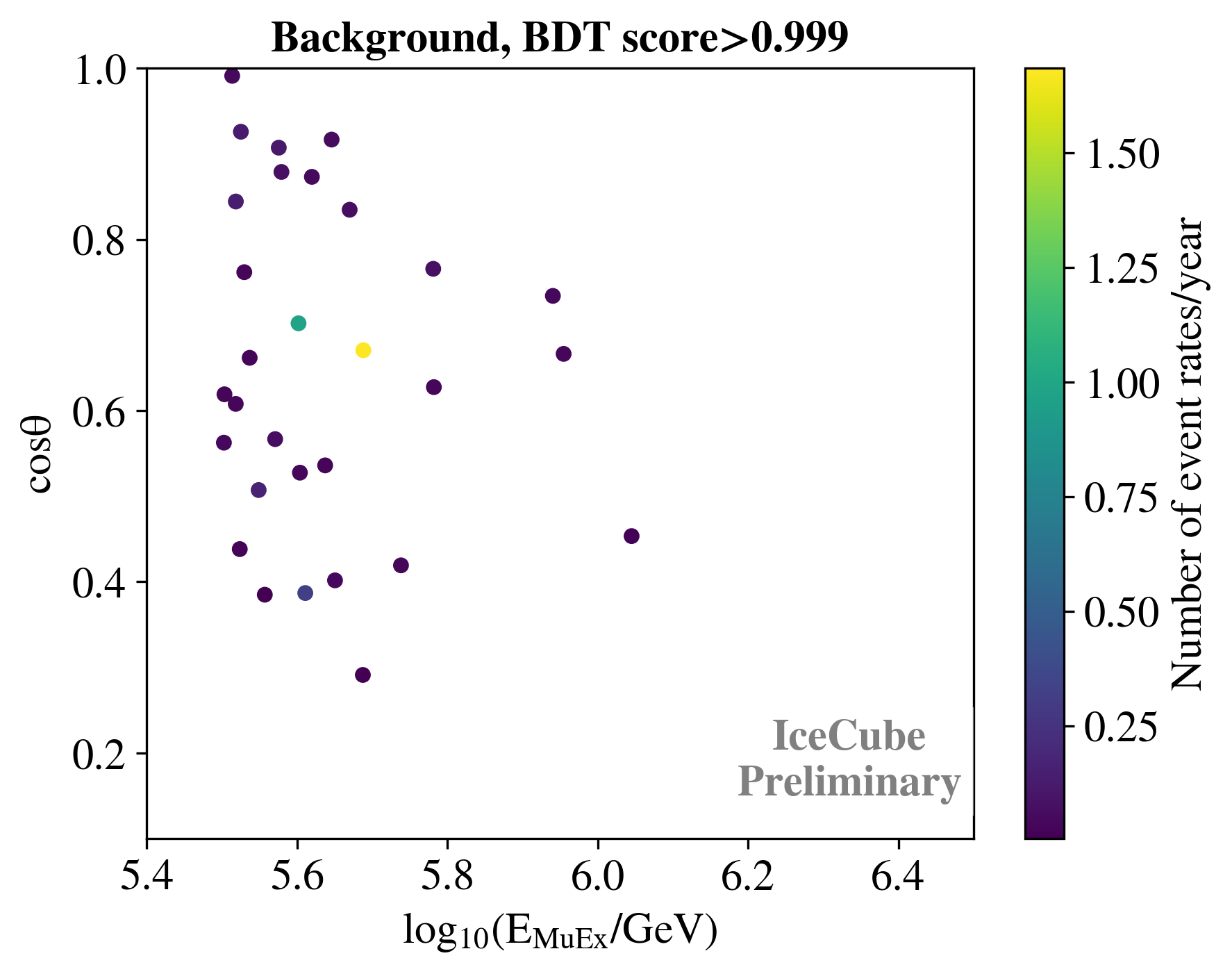}
  \end{subfigure}
  \caption{First two plots show distributions of BDT score for cosmic-ray (red) and neutrino (blue) events. The first plot shows a clear separation between the two classes. The plot in the middle provides a zoomed-in view with finer binning, focusing on the high-score region where signal-like (neutrino) events dominate. Zenith and energy distribution of cosmic rays with BDT score$~>0.999$ is shown in the last plot. Here, the passing backgrounds are mostly low-energy near-vertical events. This is consistent with the high rate of cosmic-ray muons entering the detector from that region of the sky.}
    \label{BDT_plot}
    \vspace{-1\baselineskip}
\end{figure}

\textbf{$\boldsymbol{\Delta\log\mathcal{L}}$:}
 The log-likelihood ratio, $\Delta\log\mathcal{L}$ is the difference between the log-likelihoods under the neutrino and cosmic-ray hypotheses for each DOM in the event. This difference is summed over all contributing DOMs within the selection range for an event. Only DOMs with time residuals in the range [$-500$, $500$]~ns and within $200$~m of the reconstructed track are included in the calculation as shown in Fig.~\ref{fig_nu_cos_2}. We use time residual, perpendicular distance to the track, total charge, and earliest charge in each DOM to construct a probability density function (PDF) for calculating the event likelihood. The PDF is built using kernel density estimation (KDE). 
\begin{figure}
\centering
\includegraphics[width=0.8\linewidth]{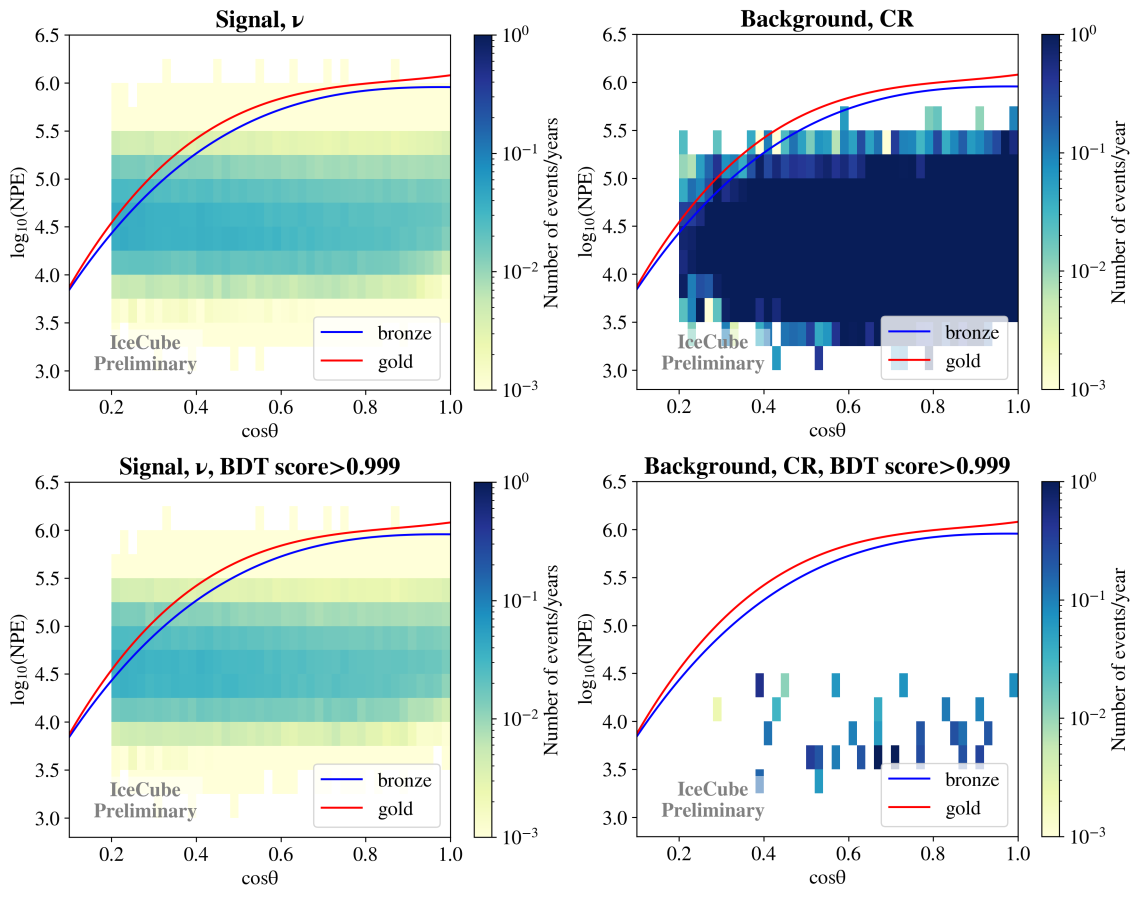}
  \caption{Plots showing GFU candidate events for both neutrino and cosmic-ray samples. The red and blue lines indicate the gold and bronze selection thresholds from~\cite{collaboration_realtime_2017}, respectively. Based on this simulation, the current alert selection yields a total of 1.34 gold events per year, including 1.1 background events. The lower plots display events with BDT score > $0.999$, effectively removing most of the background while keeping the majority of signal-like events.}
  \label{alertplots}
  \vspace{-5pt}
\end{figure}

\begin{figure}[t]
  \begin{subfigure}[b]{0.32\textwidth}
    \includegraphics[width=\textwidth]{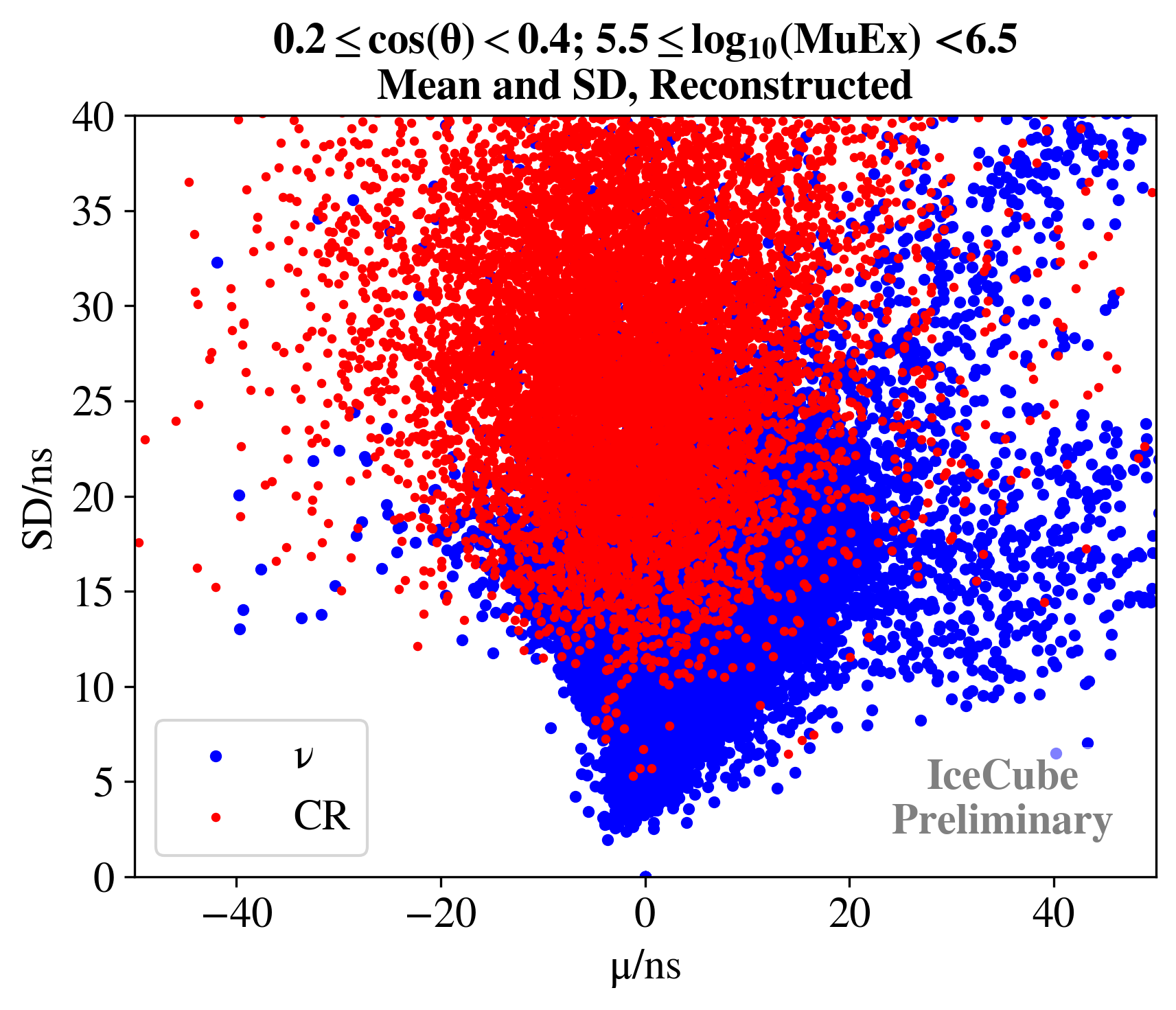}
  \end{subfigure}
  \hfill
  \begin{subfigure}[b]{0.32\textwidth}
    \includegraphics[width=\textwidth]{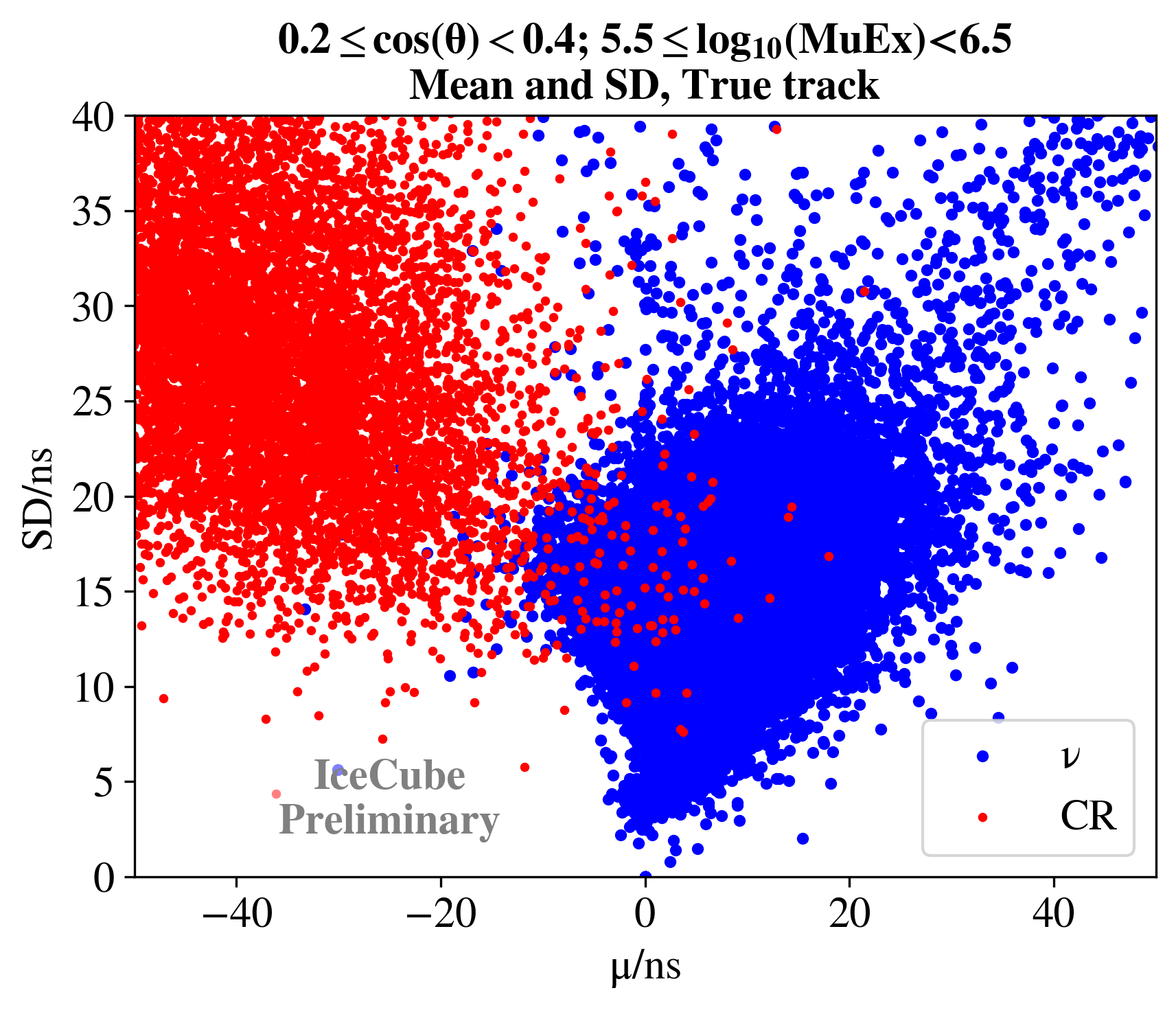}
  \end{subfigure}
  \hfill
  \begin{subfigure}[b]{0.32\textwidth}
    \includegraphics[width=\textwidth]{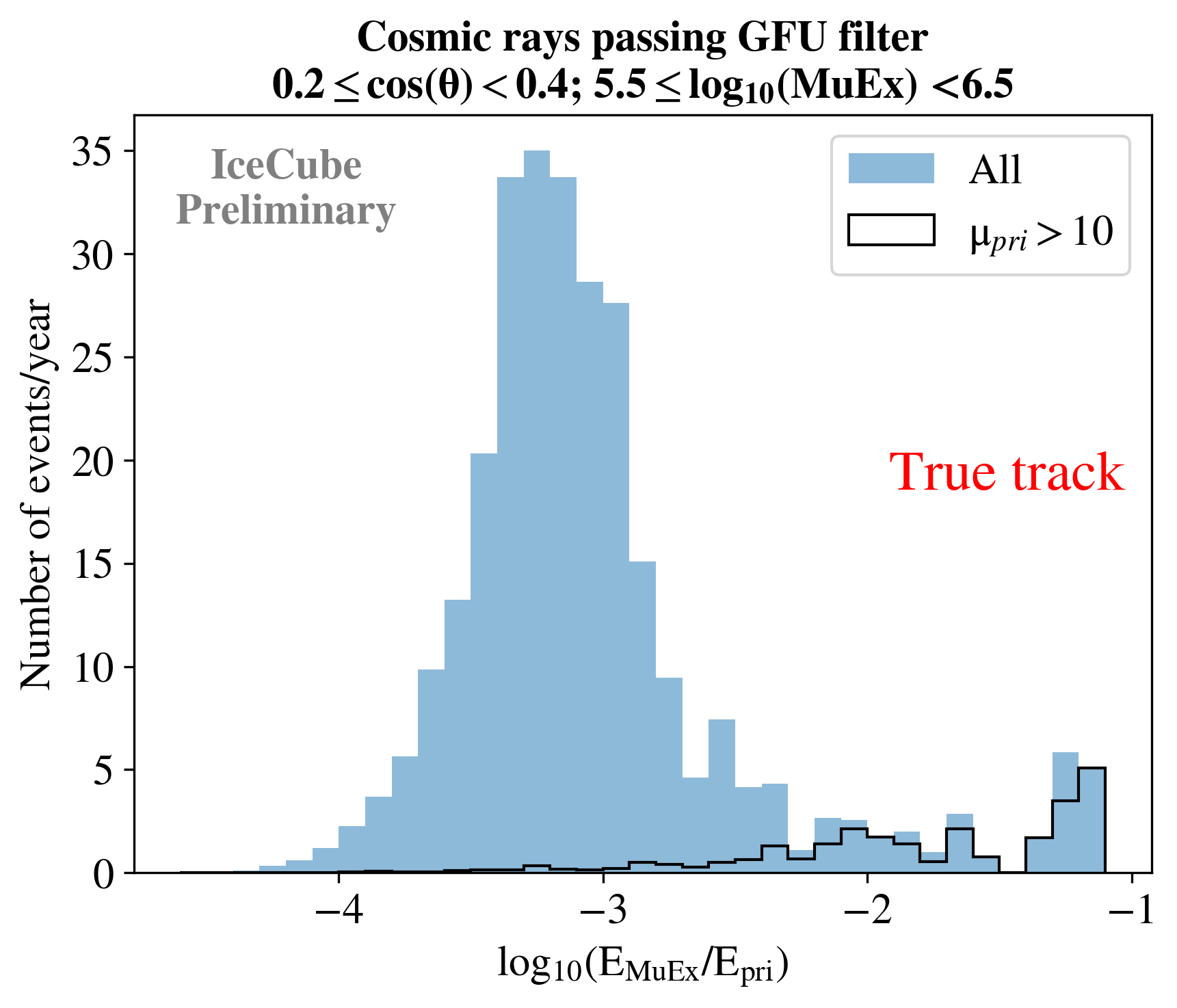}
  \end{subfigure}
  \caption{Scatter plots of mean and SD showing the distribution of neutrino and cosmic-ray events using reconstructed and true track information. The presence of misclassified cosmic-ray events persists even with perfect reconstruction. These events are primarily caused by leading muons, as illustrated in the final panel.}
  \label{primary}
\end{figure}

\textbf{Mean, Standard Deviation and Skewness:}
We calculate the mean and standard deviation (SD) of time residual of the pulses within the selection range specified in section~\ref{sec2}. The early pulses from parallel muons result in a lower mean time residual and higher SD for cosmic-ray events compared to neutrinos. Moreover, the track reconstruction assumes that all the detected light comes from a single muon. This can misrepresent muon bundle reconstruction by ignoring their additional light as shown in Fig.~\ref{primary}. To capture timing inconsistencies from using the single-muon hypothesis on bundles, we use skewness, which quantifies asymmetry in the time residual distribution around its mean. We found that the skewness of DOMs with time residuals less than $150$~ns and located within $100$~m from the track shows a better separation between neutrino events and cosmic rays.

\section{BDT Performance and Results}\label{result}

To distinguish neutrino events from the dominant cosmic-ray background, we use a BDT trained on the variables described in section~\ref{sec3}. Higher BDT scores indicate more signal-like events. Based on simulations, applying a threshold of BDT score > $0.999$ selects about $4$ neutrino events per year, along with approximately $4.4$ cosmic-ray background events. By tightening the cut to BDT $>0.9999$, the background is reduced to just $0.26$ cosmic-ray events per year, while still keeping about $\sim3$ neutrino events. This allows for a high-purity selection of astrophysical neutrino candidates.
The classifier performs best for inclined events because of the greater ice overburden that enhances neutrino interaction probability and higher detector resolution due to denser DOM spacing. Most of the cosmic-ray events passing the BDT score > $0.999$ cut originate from comparatively low-energy, vertical events, many of which can be tagged by IceTop~\cite{amin_implementation_2021}.

Fig.~\ref{alertplots} shows GFU candidate events along with the alert cuts. An event is issued as an alert if it passes a two-dimensional selection based on the reconstructed zenith angle and the deposited charge, quantified by the number of photoelectrons (NPE), as described in~\cite{collaboration_realtime_2017}. This zenith-dependent charge cut effectively suppresses a large portion of the background at the cost of a significant loss of signal. In contrast, applying a BDT score > 0.999 significantly reduces the background while maintaining high signal purity.  

\section{Conclusion}\label{conclusion} 
We presented a BDT-based approach to separate neutrino candidates from cosmic-ray backgrounds using topological and charge-based variables. Although the simulations used in this analysis do not account for the birefringent properties ~\cite{abbasi_situ_2024}  of South Pole ice, we do not expect a significant impact as the analysis focuses on DOMs close to the track where the effect will be minimal. The performance of the analysis is strongly dependent on the quality of event reconstruction. Fig.~\ref{primary} demonstrates that the overall performance of the classifier will improve with better reconstruction. The irreducible background is primarily composed of leading atmospheric muons that penetrate deep into the detector and can closely resemble signal events.  The method achieves high signal purity while effectively reducing background.  It has the potential to improve performance over traditional cuts and enhance the quality of neutrino alert selection at PeV energies and above.

\bibliographystyle{ICRC}
\vspace{-0.5cm}

\begin{thebibliography}{10}

\bibitem{collaboration_icecube}
{\bfseries IceCube} Collaboration, M.~Aartsen {\em et~al.},
  \href{http://dx.doi.org/10.1088/1748-0221/12/03/P03012}{{\em JINST}
  {\bfseries 12} no.~03, (2017) }.

\bibitem{collaboration_icetop_2013}
{\bfseries IceCube} Collaboration, M.~Aartsen {\em et~al.},
  \href{http://dx.doi.org/10.1016/j.nima.2012.10.067}{{\em NIM-A} {\bfseries
  700} }.

\bibitem{collaboration_calibration_2010}
{\bfseries IceCube} Collaboration, R.~Abbasi {\em et~al.},
  \href{http://dx.doi.org/10.1016/j.nima.2010.03.102}{{\em NIM-A} {\bfseries
  618} no.~1-3, (2010) }.

\bibitem{science_2013}
{\bfseries IceCube} Collaboration, M.~Aartsen {\em et~al.},
  \href{http://dx.doi.org/10.1126/science.1242856}{{\em Science} (2013) }.

\bibitem{collaboration_realtime_2017}
{\bfseries IceCube} Collaboration, M.~Aartsen {\em et~al.},
  \href{http://dx.doi.org/10.1016/j.astropartphys.2017.05.002}{{\em
  Astroparticle Physics} {\bfseries 92} (2017) }.

\bibitem{heck_extensive_2001}
D.~Heck, 2001.
\newblock \url{http://arxiv.org/abs/astro-ph/0103073}.

\bibitem{Sybill2.3}
F.~Riehn {\em et~al.}
\newblock 2016.
\newblock \url{https://pos.sissa.it/236/558}.

\bibitem{GaisserH4A}
T.~K. {Gaisser},
  \href{http://dx.doi.org/10.1016/j.astropartphys.2012.02.010}{{\em
  Astroparticle Physics} {\bfseries 35} no.~12, (2012) }.

\bibitem{nugen}
A.~Gazizov {\em et~al.}, 2005.
\newblock \url{https://doi.org/10.1016/j.cpc.2005.03.113}.

\bibitem{abbasi_improved_2022}
{\bfseries IceCube} Collaboration, R.~Abbasi {\em et~al.},
  \href{http://dx.doi.org/10.3847/1538-4357/ac4d29}{{\em The Astrophysical
  Journal} {\bfseries 928} no.~1, (2022) }.

\bibitem{rongen_calibration_nodate}
M.~Rongen Master's thesis, 2019.
\newblock \url{http://arxiv.org/abs/1911.02016}.

\bibitem{kintscher_rapid_2020}
T.~Kintscher.
\newblock PhD thesis, 2020.
\newblock \url{https://doi.org/10.18452/21948}.

\bibitem{collaboration_energy_2014}
{\bfseries IceCube} Collaboration, M.~Aartsen {\em et~al.},
  \href{http://dx.doi.org/10.1088/1748-0221/9/03/P03009}{{\em JINST} {\bfseries
  9} no.~03, (2014) }.

\bibitem{schatto_stacked_2014}
K.~Schatto.
\newblock PhD thesis, June, 2014.
\newblock \url{https://inspirehep.net/literature/1409209}.

\bibitem{yang_probing_2025}
{\bfseries IceCube} Collaboration, R.~Abbasi {\em et~al.},
  \href{http://dx.doi.org/10.1103/2hnq-1fsx}{{\em Physical Review D} (2025) }.

\bibitem{amin_implementation_2021}
{\bfseries IceCube} Collaboration, N.~M.~B. Amin,
  \href{http://dx.doi.org/10.1088/1742-6596/2156/1/012217}{{\em JPCS}
  {\bfseries 2156} no.~1, (2021) }.

\bibitem{abbasi_situ_2024}
{\bfseries IceCube} Collaboration, R.~Abbasi {\em et~al.},
  \href{http://dx.doi.org/10.5194/tc-18-75-2024}{{\em The Cryosphere}
  {\bfseries 18} no.~1, (2024) }.

\end{thebibliography}
\providecommand{\href}[2]{#2}\begingroup\raggedright\endgroup

\clearpage

\section*{Full Author List: IceCube Collaboration}

\scriptsize
\noindent
R. Abbasi$^{16}$,
M. Ackermann$^{63}$,
J. Adams$^{17}$,
S. K. Agarwalla$^{39,\: {\rm a}}$,
J. A. Aguilar$^{10}$,
M. Ahlers$^{21}$,
J.M. Alameddine$^{22}$,
S. Ali$^{35}$,
N. M. Amin$^{43}$,
K. Andeen$^{41}$
% ...
\\
\\
% ...
$^{10}$ Universit{\'e} Libre de Bruxelles, Science Faculty CP230, B-1050 Brussels, Belgium \\
% ...
$^{16}$ Department of Physics, Loyola University Chicago, Chicago, IL 60660, USA \\
$^{17}$ Dept. of Physics and Astronomy, University of Canterbury, Private Bag 4800, Christchurch, New Zealand \\
% ...
$^{21}$ Niels Bohr Institute, University of Copenhagen, DK-2100 Copenhagen, Denmark \\
$^{22}$ Dept. of Physics, TU Dortmund University, D-44221 Dortmund, Germany \\
% ...
$^{35}$ Dept. of Physics and Astronomy, University of Kansas, Lawrence, KS 66045, USA \\
% ...
$^{39}$ Dept. of Physics and Wisconsin IceCube Particle Astrophysics Center, University of Wisconsin{\textemdash}Madison, Madison, WI 53706, USA \\
% ...
$^{41}$ Department of Physics, Marquette University, Milwaukee, WI 53201, USA \\
% ...
$^{43}$ Bartol Research Institute and Dept. of Physics and Astronomy, University of Delaware, Newark, DE 19716, USA \\
% ...
$^{63}$ Deutsches Elektronen-Synchrotron DESY, Platanenallee 6, D-15738 Zeuthen, Germany \\
$^{\rm a}$ also at Institute of Physics, Sachivalaya Marg, Sainik School Post, Bhubaneswar 751005, India \\
% ...

\subsection*{Acknowledgments}

\noindent
The authors gratefully acknowledge the support from the following agencies and institutions:
USA {\textendash} U.S. National Science Foundation-Office of Polar Programs,
U.S. National Science Foundation-Physics Division,
U.S. National Science Foundation-EPSCoR,
U.S. National Science Foundation-Office of Advanced Cyberinfrastructure,
Wisconsin Alumni Research Foundation,
Center for High Throughput Computing (CHTC) at the University of Wisconsin{\textendash}Madison,
Open Science Grid (OSG),
Partnership to Advance Throughput Computing (PATh),
Advanced Cyberinfrastructure Coordination Ecosystem: Services {\&} Support (ACCESS),
Frontera and Ranch computing project at the Texas Advanced Computing Center,
U.S. Department of Energy-National Energy Research Scientific Computing Center,
Particle astrophysics research computing center at the University of Maryland,
Institute for Cyber-Enabled Research at Michigan State University,
Astroparticle physics computational facility at Marquette University,
NVIDIA Corporation,
and Google Cloud Platform;
Belgium {\textendash} Funds for Scientific Research (FRS-FNRS and FWO),
FWO Odysseus and Big Science programmes,
and Belgian Federal Science Policy Office (Belspo);
Germany {\textendash} Bundesministerium f{\"u}r Forschung, Technologie und Raumfahrt (BMFTR),
Deutsche Forschungsgemeinschaft (DFG),
Helmholtz Alliance for Astroparticle Physics (HAP),
Initiative and Networking Fund of the Helmholtz Association,
Deutsches Elektronen Synchrotron (DESY),
and High Performance Computing cluster of the RWTH Aachen;
Sweden {\textendash} Swedish Research Council,
Swedish Polar Research Secretariat,
Swedish National Infrastructure for Computing (SNIC),
and Knut and Alice Wallenberg Foundation;
European Union {\textendash} EGI Advanced Computing for research;
Australia {\textendash} Australian Research Council;
Canada {\textendash} Natural Sciences and Engineering Research Council of Canada,
Calcul Qu{\'e}bec, Compute Ontario, Canada Foundation for Innovation, WestGrid, and Digital Research Alliance of Canada;
Denmark {\textendash} Villum Fonden, Carlsberg Foundation, and European Commission;
New Zealand {\textendash} Marsden Fund;
Japan {\textendash} Japan Society for Promotion of Science (JSPS)
and Institute for Global Prominent Research (IGPR) of Chiba University;
Korea {\textendash} National Research Foundation of Korea (NRF);
Switzerland {\textendash} Swiss National Science Foundation (SNSF).

\end{document}